\documentstyle[12pt,epsf,epsfig,cite]{article}

\newcommand{\nc}{\newcommand}
\nc{\postscript}[2]
{\setlength{\epsfxsize}{#2\hsize}\centerline{\epsfbox{#1}}}
\nc{\bg}{B. Grzadkowski}
\nc{\non}{\nonumber}
\nc{\hc}{\hbox {h.c.}} \nc{\re}{\hbox {Re}} 
\nc{\mev}{\hbox {MeV}} \nc{\gev}{\;\hbox {GeV}} \nc{\tev}{\;\hbox {TeV}}
\def\lsim{\mathrel{\raise.3ex\hbox{$<$\kern-.75em\lower1ex\hbox{$\sim$}}}}
\def\gsim{\mathrel{\raise.3ex\hbox{$>$\kern-.75em\lower1ex\hbox{$\sim$}}}}

\nc{\prd}[3]{{\it Phys.\ Rev.}\ {{\bf D{#1}} (#2), #3}}
\nc{\prl}[3]{{\it Phys.\ Rev.\ Lett.}\ {{\bf {#1}} (#2), #3}}
\nc{\plb}[3]{{\it Phys.\ Lett.}\ {{\bf B{#1}} (#2), #3}}
\nc{\npb}[3]{{\it Nucl.\ Phys.}\ {{\bf B{#1}} (#2), #3}}
\nc{\ptp}[3]{{\it Prog.\ Theor.\ Phys.}\ {{\bf {#1}} (#2), #3}}
\nc{\zfp}[3]{{\it Z.\ Phys.}\ {{\bf C{#1}} (#2), #3}}
\nc{\epj}[3]{{\it Eur.\ Phys.\ J.}\ {{\bf C{#1}} (#2), #3}}
\nc{\mpla}[3]{{\it Mod.\ Phys.\ Lett.}\ {{\bf A{#1}} (#2), #3}}
\nc{\rmp}[3]{{\it Rev.\ Mod.\ Phys.}\ {{\bf {#1}} (#2), #3}}
\nc{\ijmpa}[3]{{\it Int.\ J.\ of\ Mod.\ Phys.}\
               {{\bf A{#1}} (#2), #3}}
\nc{\Lsp}{\;\;\;\;\;\;\;\;\;\;}  \nc{\LLLsp}{\lspace \lspace}
\nc{\lsp}{\;\;\;\;\;\;}
\nc{\spac}{\;\;\;}
\nc{\noi}{\noindent}
\nc{\beq}{\begin{equation}}   \nc{\eeq}{\end{equation}}
\nc{\bea}{\begin{eqnarray}}   \nc{\eea}{\end{eqnarray}}
\nc{\baa}{\begin{array}}      \nc{\eaa}{\end{array}}
\nc{\bit}{\begin{itemize}}    \nc{\eit}{\end{itemize}}
\nc{\ben}{\begin{enumerate}}  \nc{\een}{\end{enumerate}}
\nc{\bce}{\begin{center}}     \nc{\ece}{\end{center}}

\def\Hhat{\widehat H}

\def\ho{h_0}
\def\mho{m_{\ho}}
\def\phio{\phi_0}
\def\mphio{m_{\phio}}

\def\anti{\overline}

\def\gam{\gamma}

\def\mh{m_{h}}
\def\mphi{m_\phi}

\def\lphi{\Lambda_\phi}

\def\mphi{m_\phi}

\def\hbar{\overline h}

\def\mpl{M_{Pl}}

\def\ifmath#1{\relax\ifmmode #1\else $#1$\fi}
\def\half{\ifmath{{\textstyle{1 \over 2}}}}

\def\call{{\cal L}}

\def\eps{\epsilon}

\begin{document}

\begin{titlepage}
\pagestyle{empty}
\baselineskip=21pt
\rightline{hep-ph/0304245}
%\rightline{CERN--TH/2003-XXX}
\rightline{ }
\vskip 0.05in
\begin{center}
{\large{\bf On the Complementarity of Higgs and Radion Searches at LHC}}
\end{center}
\begin{center}
\vskip 0.08in
{{\bf Marco~Battaglia}$^1$,
{\bf Stefania De Curtis}$^2$,
{\bf Albert De Roeck}$^1$, \\
{\bf Daniele Dominici}$^{2,3}$ and
{\bf John F. Gunion}$^4$
}
\vskip 0.08in
{\it
$^1${CERN, Geneva (Switzerland)}\\
$^2${INFN, Sezione di Firenze (Italy)}\\
$^3${Universita' degli Studi di Firenze, Dip. di Fisica (Italy)}\\
$^4${University of California, Davis CA (USA)}
}
\vskip 0.35in
{\bf Abstract}
\end{center}
\baselineskip=18pt \noindent
%%%%%%%%%%%%%%%%%%%%%%%%%%%%%%%%%%%%%%%%%%%%%%%%%%%%%%%%%%%%%%%%%%%%%

Models with 3-branes in extra dimensions typically
imply the existence of a radion, $\phi$, that can mix with the Higgs, $h$,
thereby modifying the Higgs properties and the prospects for 
its detectability at the LHC. The presence of the $\phi$ will extend
the scope of the LHC searches. Detection of both the $\phi$ and the $h$
might be possible. In this paper, we study the complementarity of the 
observation of $gg \to h$, with $h \to \gamma \gamma$
or $h\to Z^0Z^{0*}\to 4~\ell$, and 
$gg \to \phi \to Z^0Z^{0(*)} \to 4~\ell$ at the LHC 
in the context of the Randall-Sundrum 
model. The potential for determining the nature of the detected scalar(s) 
at the LHC and at an $e^+e^-$ linear collider is discussed, 
both separately and in combination.

%%%%%%%%%%%%%%%%%%%%%%%%%%%%%%%%%%%%%%%%%%%%%%%%%%%%%%%%%%%%%%%%%%%%%
\vfill
\vskip 0.15in
%\leftline{CERN--TH/2003-XXX}
\leftline{25 April 2003}
\end{titlepage}
\baselineskip=18pt
%%%%%%%%%%%%%%%%%%%%%%%%%%%%%%%%%%%%%%%%%%%%%%%%%%%%%%%%%%%%%%%%%%%%%

\section{Introduction}

By the end of this decade we expect that the 
quest for the Higgs boson, responsible for 
electro-weak symmetry breaking and mass generation, 
will be successfully completed, thanks
to the data provided by the LHC hadron collider. 
A significant effort has been put into the
design and optimization of the {\sc Atlas} 
and {\sc Cms} detectors to match the 
characteristics of the expected Higgs signals. However fundamental, 
discovery of one or more Higgs-like particles might leave unanswered 
the question of the hierarchy between the electroweak scale, defined by
the Higgs vacuum expectation value $v=$246~GeV, 
and the Planck scale. In an attempt to 
solve this problem, without necessarily relying on supersymmetry, 
theories with extra dimensions have been proposed. These theories 
have become the focus of a fascinating program of planned investigations.

One particularly attractive extra-dimensional model is that proposed
by Randall and Sundrum (RS)~\cite{rs}, in which there are two 3+1 dimensional
branes separated in a 5th dimension. A central prediction of
this theory is the existence of the radion,  a graviscalar which 
corresponds to fluctuations in the size of the extra 
dimension. Detection and study of the radion
will be central to the experimental probe of the RS and related
scenarios with extra dimensions. 
There is already an extensive literature on the phenomenology of the radion,
both in the absence of Higgs-radion 
mixing~\cite{Bae:2000pk,Davoudiasl:1999jd,Cheung:2000rw,Davoudiasl:2000wi,Park:2000xp}
and in the presence of such a mixing~\cite{wells_mix,csaki_mix,Han:2001xs,Chaichian:2001rq,
Azuelos:fv,Hewett:2002nk,Dominici:2002jv}.

In this paper we discuss the complementarity of the search for the Higgs boson and the 
radion at the LHC. As the Higgs-radion mixing may suppress the main discovery process 
$gg \to H \to \gamma \gamma$ for a light Higgs boson, we study the extent
to which  the 
appearance of a $gg \to \phi \to Z^0Z^{0(*)} \to 4~\ell$ signal ensures that LHC 
experiments will observe at least one of the two scalars over the full parameter phase 
space. The additional information, which could be extracted from a TeV-class 
$e^+e^-$ linear collider (LC), is also considered.

\section{Curvature-Scalar mixing and Radion Phenomenology}
\label{secmixing}

In the simplest version of the 5-dimensional RS model, all the SM particles and forces, 
with the exception of gravity, are confined to one of the 4-dimensional boundaries. 
Gravity lives on this visible brane, on the second hidden brane and in the 5-dimensional
compactified bulk. All mass scales in the 5-dimensional theory are of the order of 
the Planck mass.  By placing the SM fields on the visible brane, all the terms of order 
of the Planck mass are rescaled by an exponential suppression factor 
$\Omega_0\equiv e^{-m_0 b_0/2}$, which is called the warp factor. This reduces the 
mass scales on the visible brane down to the weak scale ${\cal O}(1 \tev)$ without any 
severe fine tuning. A ratio of $1\tev/\mpl$ (where $\mpl$ is the reduced Planck mass, 
$\mpl\sim 2.4\times 10^{18}\gev$) corresponds to $m_0 b_0 /2\sim 35$. 

In the RS model, a mixing between the radion field and the Higgs field 
$\Hhat$ is induced by the following action~\cite{johum}: 
\beq S_\xi=\xi \int d^4 x
\sqrt{g_{\rm vis}}R(g_{\rm vis})\Hhat^\dagger \Hhat\,, 
\eeq 
where $R(g_{\rm vis})$ is the Ricci scalar for the metric induced on the
visible brane, $g^{\mu\nu}_{\rm
vis}=\Omega_0^2 \Omega^2(x)(\eta^{\mu\nu}+\eps h^{\mu\nu})$ and $\xi$
a dimensionless parameter.
 After rescaling
$H_0=\Omega_0 \Hhat$,  and making the usual shifts
[$ H_0={1\over \sqrt 2}(v+\ho)\,,\ \ 
\Omega(x)=1+{\phi_0\over \lphi}\, $,  with $v=246\gev$]
 the following kinetic energy terms are found:
 \beq
\call=-\half\left\{1+6\gamma^2 \xi \right\}\phi_0\Box\phi_0
-\half\phi_0 \mphio^2\phi_0-\half h_0 (\Box+\mho^2)h_0-6\gamma \xi
\phi_0\Box h_0\,, \label{keform} \eeq where
$\mho$ and $\mphio$ are the Higgs and
radion masses before mixing, and $\gamma\equiv v/\lphi$.

The states that diagonalize the kinetic energy and
have canonical normalization are $h$ and $\phi$ with: 
\bea
h_0&=&\left (\cos\theta -{6\xi\gam\over Z}\sin\theta\right)h
+\left(\sin\theta+{6\xi\gam\over Z}\cos\theta\right)\phi\equiv d
h+c\phi
\label{hform}\\
\phi_0&=&-\cos\theta {\phi\over Z}+\sin\theta {h\over Z}\equiv a\phi+bh\,. \label{phiform}
\eea
Here, the mixing angle $\theta$ is given by 
\beq \tan 2\theta\equiv 12 \gam \xi Z
{\mho^2\over \mphio^2-\mho^2(Z^2-36\xi^2\gam^2)}\,, \label{theta}
\eeq 
and
\beq Z^2\equiv 1+6\xi\gam^2(1-6\xi)\,. \label{z2} \eeq 
The couplings of the $h$ and $\phi$ to $ZZ$, $WW$ and $f\anti f$ are given
relative to those of the SM Higgs boson, denoted by $H$, by:
\beq
{g_{hW^+W^-}\over g_{HW^+W^-}}={g_{hZ^0Z^0}\over g_{HZ^0Z^0}}={g_{hf\anti f}\over g_{H f\anti f}}=d+\gam b\,,\quad
{g_{\phi W^+W^-}\over g_{HW^+W^-}}={g_{\phi Z^0Z^0}\over g_{HZ^0Z^0}}={g_{\phi f\anti f}\over g_{H f\anti f}}=c+\gam a\,. 
\label{couplings}
\eeq
Couplings of the $h$ and $\phi$ to $\gamma\gamma$ and $gg$ 
receive contributions not only from the usual loop diagrams but also
from trace-anomaly couplings of the $\phi_0$ to $\gamma\gamma$ and $gg$.
Thus, these couplings are not simply directly proportional to
those of the SM $H$.
Of course, 
in the limit of $\xi=0$, the $h$ has the same properties as the SM Higgs 
boson.

In the end, 
when $\xi\neq0$ the four primary independent parameters 
are: $\xi$, $\lphi$ and the mass eigenvalues $\mh$ and $\mphi$.
These completely determine $a,b,c,d$ and, hence,
all the couplings of the $h$ and $\phi$
to $W^+W^-$, $Z^0Z^0$ and $f\anti f$ --- see Eq.~(\ref{couplings}). 
One further parameter is required to completely fix
the $h$ and $\phi$ decay phenomenology in the RS model: 
 $m_1$, the mass of the first KK graviton excitation, given by
\beq
m_1=x_1 {m_0\over\mpl} {\lphi\over\sqrt 6}\,
\label{m1form}
\eeq
where  ${m_0}$ is the curvature parameter and  $x_1$ is the first
zero of the Bessel function $J_1$ ($x_1\sim 3.8$).

Current bounds, derived from Tevatron Run~I data and precision electroweak
constraints have been examined in Ref.~\cite{Davoudiasl:2000wi}.
Lower bounds on the radion mass, from Higgs searches at LEP, are weak. 
In particular, the $\phi Z^0Z^0$ coupling given in
Eq.~(\ref{couplings}) remains
small relative to the SM $HZ^0Z^0$ coupling 
for low radion masses ~\cite{Dominici:2002jv}.

\section{Higgs and Radion Searches at LHC}

\begin{figure}[h!]
\begin{center}
\begin{tabular}{c c}
\epsfig{file=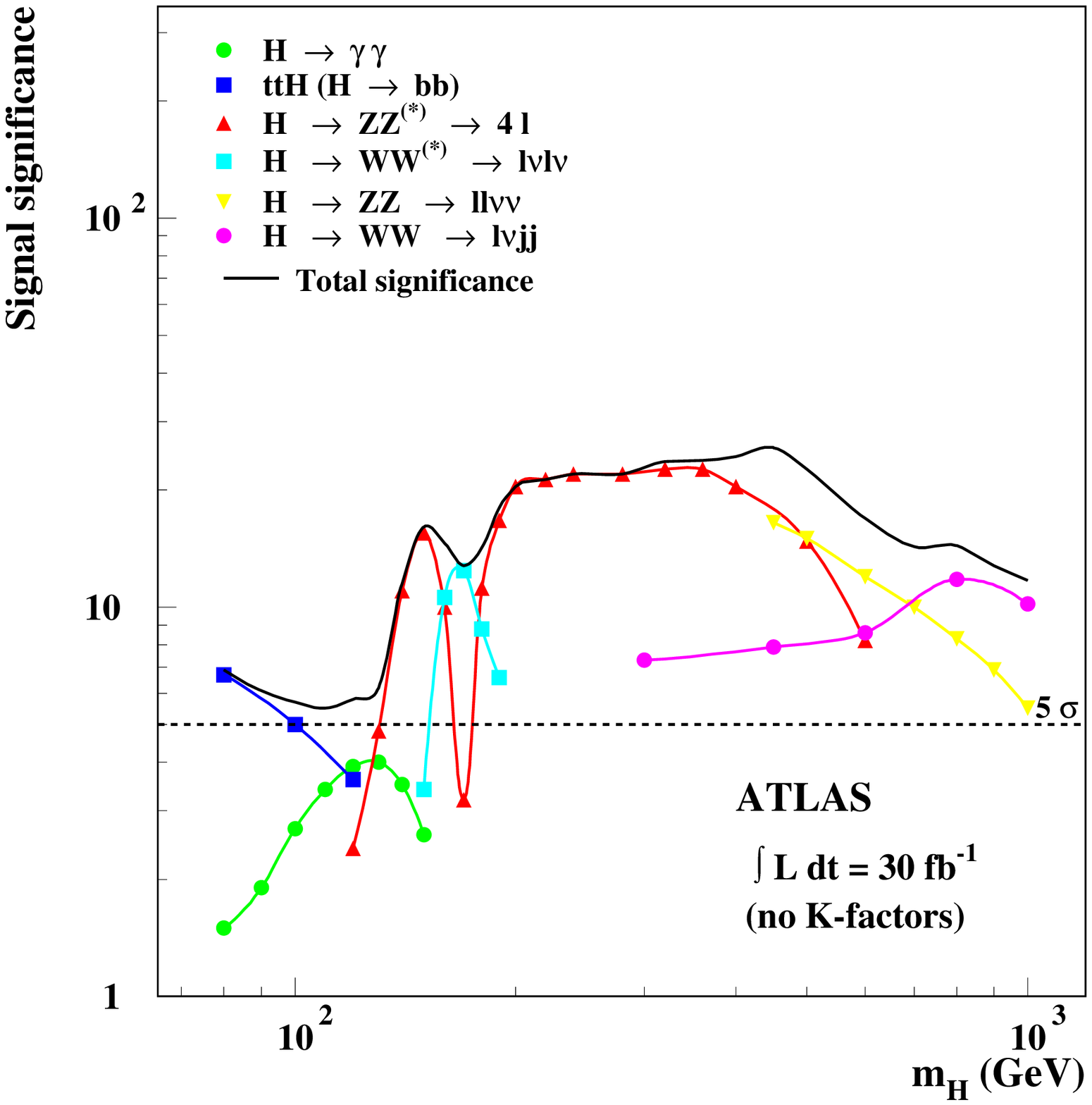,width=8.0cm,height=6.0cm} &
\epsfig{file=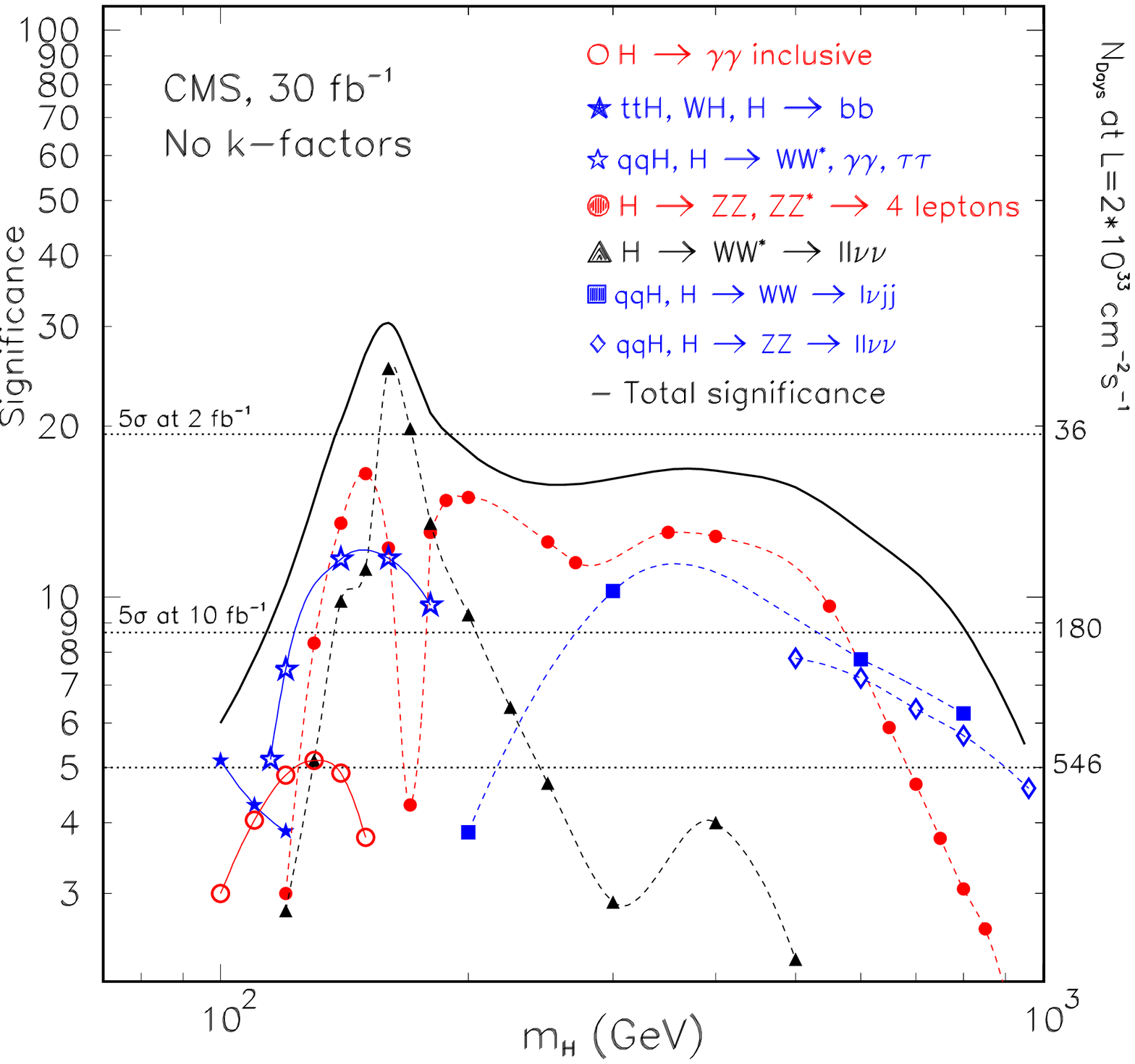,width=8.0cm,height=6.0cm} \\
\end{tabular}
\end{center}
\vspace*{-0.75cm}
\caption{\sl Higgs signal significance as function of the Higgs boson mass.
The curves show  the signal significance for an integrated luminosity of 30~fb$^{-1}$ 
for {\sc Atlas}~\cite{atlas-tdr} (left) and {\sc Cms}~\cite{cmsnote} (right). In the 
right plot the contributions from the $qqH$ channel are also shown. No K-factors have 
been included.}
\label{fig:sign}
\end{figure}

The search for the Higgs boson represents one of the most crucial goals for the 
LHC physics program. If the SM $H$
is light, as present precision electroweak 
data suggest, the single most promising LHC discovery channel is $gg \to H \to 
\gamma \gamma$. Rather detailed studies of the significance of a Higgs signal 
using inclusive production have been carried out for the {\sc Atlas}~\cite{atlas-tdr} 
and {\sc Cms}~\cite{cmsnote} experiments. Results are summarized in 
Figure~\ref{fig:sign}. The $H \to \gamma \gamma$ channel appears to be instrumental for 
obtaining a $\ge 5 \sigma$ signal at low luminosity, if 115~GeV $< M_H <$ 130~GeV. 
The $t \bar{t} H$, $H \to b \bar{b}$ and $H \to Z^0Z^{0*} \to 4~\ell$ 
channels also 
contribute, with lower statistics but a more favourable 
signal-to-background ratio. Preliminary results, also shown in Figure~\ref{fig:sign}, 
indicate that Higgs boson production in association with forward jets may also be 
considered as a discovery mode. However, here the background suppression strongly 
relies on the detailed detector response.

Studies for dedicated searches of radions at the 
LHC have also been carried out. 
In particular, the study in ref.~\cite{Azuelos:fv} has obtained discovery limits using 
both the $\phi \to \gamma \gamma$ and the $\phi \to Z^0Z^{0(*)} \to 4 \ell$ processes by 
re-interpreting the corresponding $H$ decay channels. The more intriguing process 
$\phi \to HH$ has also been considered, limited to the $\gamma \gamma b \bar{b}$ final 
state.

A more subtle aspect of theories with warped extra dimensions is the effect of the 
Higgs-radion mixing~\cite{Chaichian:2001rq,Hewett:2002nk,Dominici:2002jv}, which can modify the 
production and decay properties of the Higgs boson to weaken, or even invalidate, the 
expectations for Higgs observability obtained so far. 

\section{Complementarity and Distinguishability}

In this section, we address two issues. The first is whether there is a complementarity 
between the Higgs observability, mostly through $gg \to h \to \gamma \gamma$, and the 
$gg \to \phi \to Z^0Z^{0(*)} \to 4~\ell$ reaction, thus offering the LHC the discovery 
of at least one of the two particles over the full parameter space. The second, and 
related, issue concerns the strategies available to understand the nature of the 
discovered particle. 

The effects of the mixing of the radion with the Higgs boson have been 
studied~\cite{Dominici:2002jv} by introducing the relevant terms in the {\sc HDecay} 
program~\cite{Djouadi:1997yw}, which computes the Higgs couplings, including higher 
order QCD corrections. Couplings and widths for the radion have also been implemented.
We consider the range 50~GeV $< M_{\phi} <$ 300~GeV, whose lower end is consistent with 
present bounds derived from LEP data. We will also focus on cases
for which $M_h$ is not very large, as possibly most consistent with
precision electroweak constraints.

Results have been obtained by comparing the product of production and decay rates to 
those expected for a light SM $H$. The LHC sensitivity has been extracted by 
rescaling the results for Higgs observability, obtained assuming SM couplings. We 
define the Higgs observability as $>5~\sigma$ excess over the SM background for the 
combination of the inclusive channels: $gg \to h \to \gamma \gamma$; $t \bar{t} h$, 
$h \to b \bar{b}$ and $h \to Z^0Z^{0(*)} \to 4\ell$, as given in the left panel of 
Figure~\ref{fig:sign}. We study the results as a function of four parameters: 
the Higgs mass $M_h$, the radion mass $M_{\phi}$, the scale $\lphi$ and the mixing 
parameter $\xi$.

\subsection{Radion and Higgs Boson Search Complementarity}

\begin{figure}[h!]
\begin{center}
\begin{tabular}{c c c}
\hspace*{-0.85cm} 
\epsfig{file=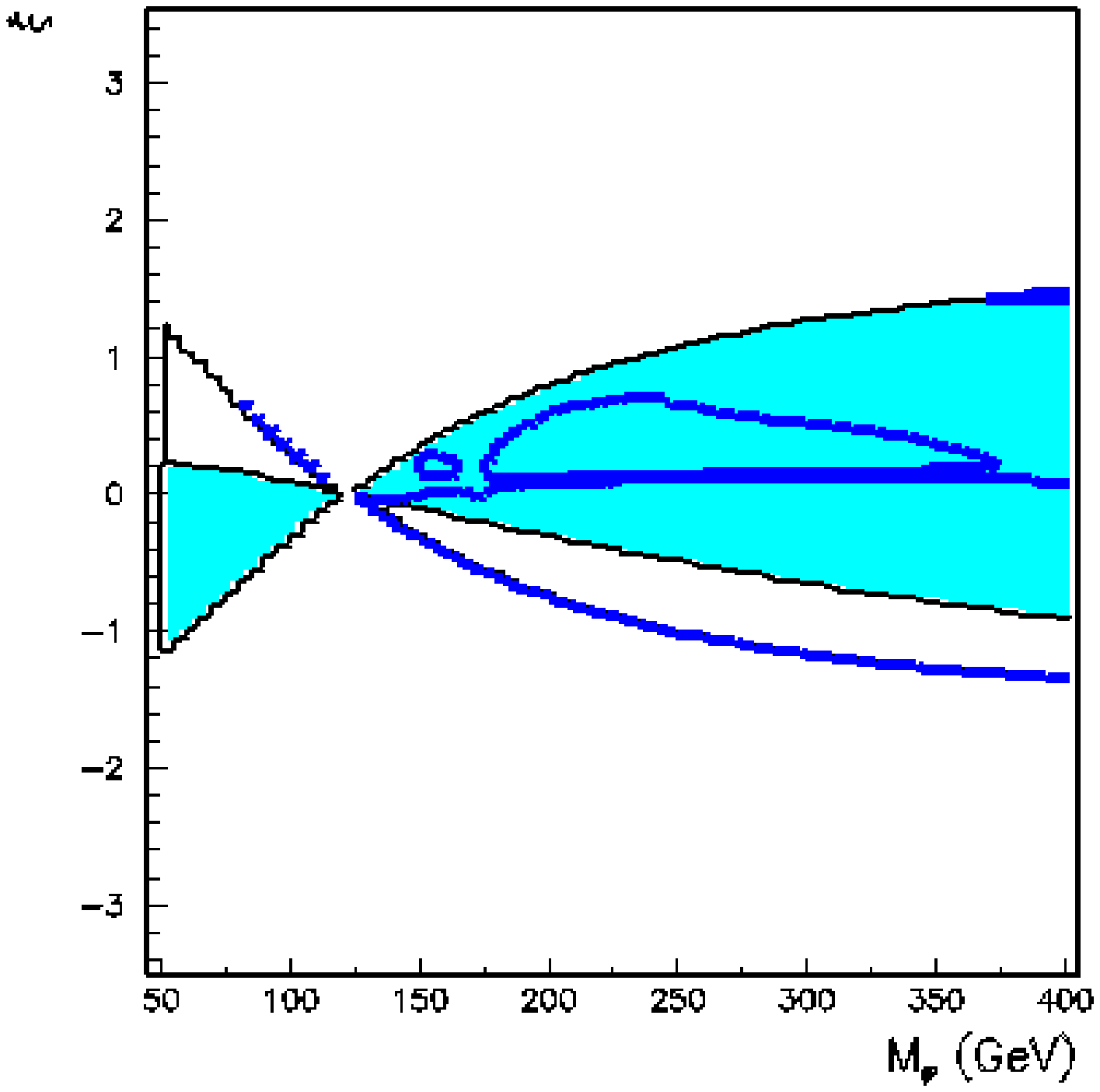,width=6.5cm,height=6.5cm} &
\hspace*{-0.7cm} 
\epsfig{file=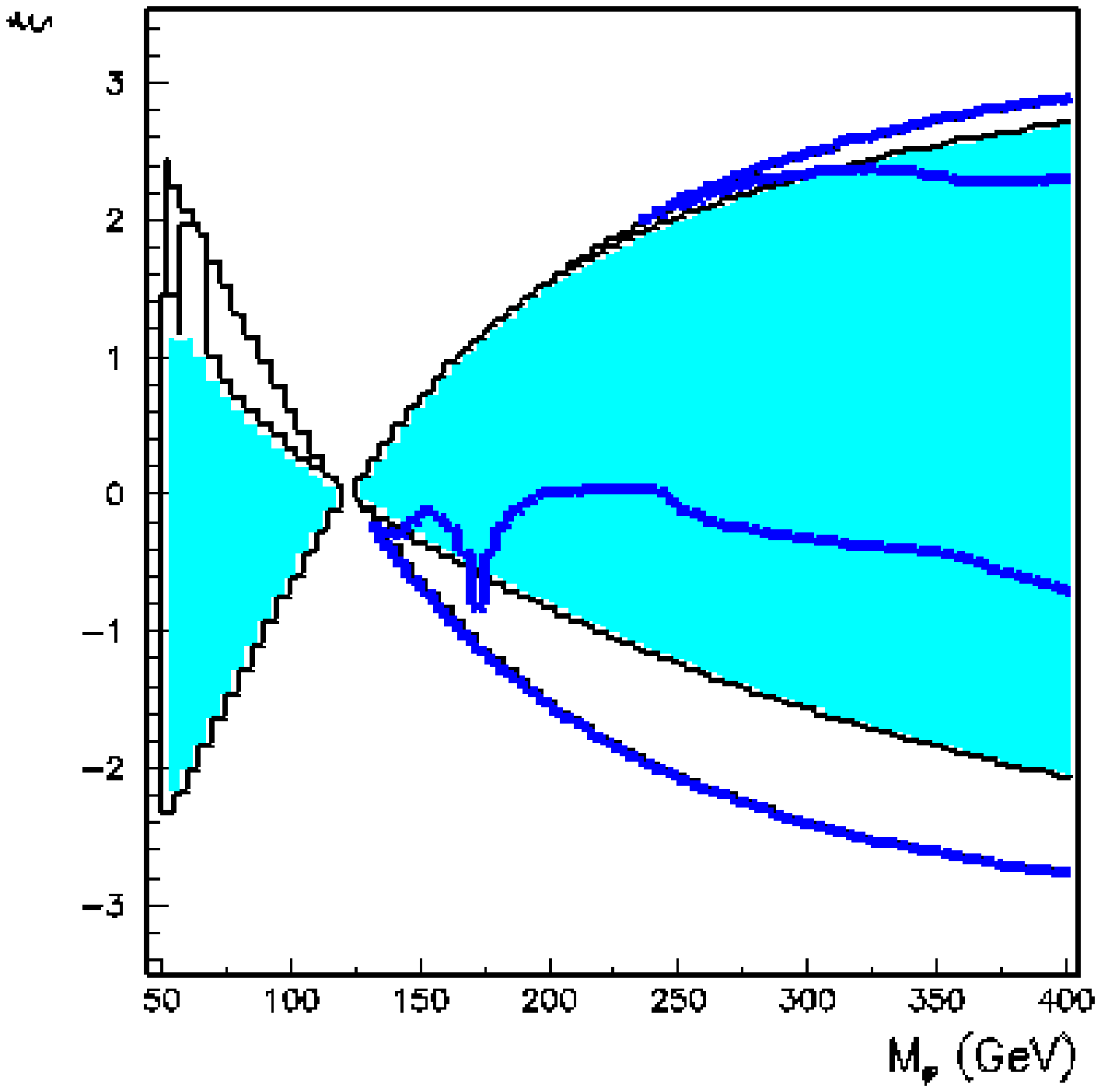,width=6.5cm,height=6.5cm} &
\hspace*{-0.7cm} 
\epsfig{file=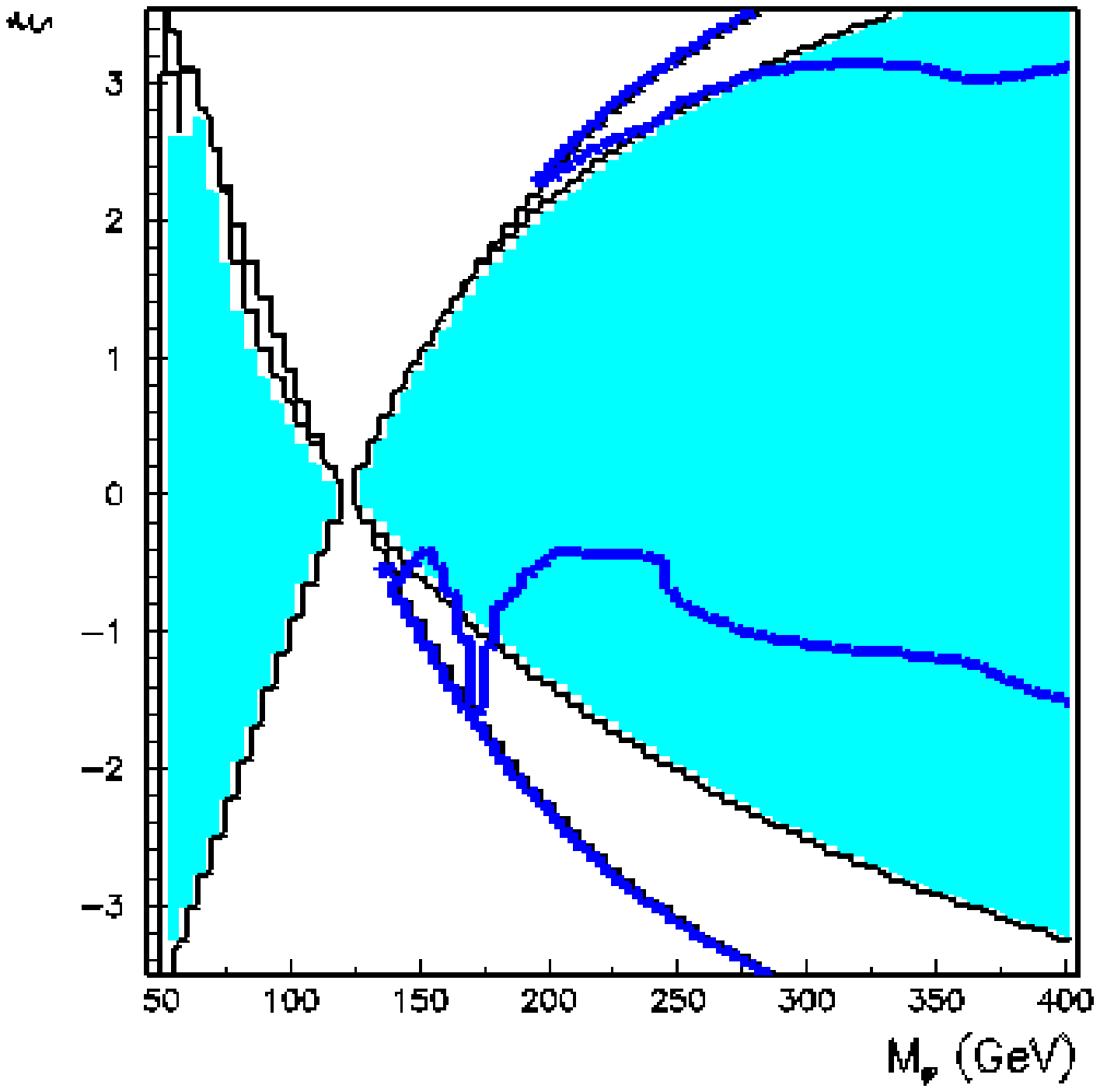,width=6.5cm,height=6.5cm} \\
\end{tabular}
\end{center}
\vspace*{-0.75cm}
\caption{\sl Regions in $(M_\phi,\xi)$ parameter space
of $h$ detectability (including $gg \to h\to \gamma\gamma$  and other modes)
and of $gg \to \phi \to Z^0Z^{0(*)} \to 4~\ell$ detectability 
at the LHC for one experiment and 30~fb$^{-1}$. 
The outermost, hourglass shaped contours 
define the theoretically allowed region. 
The light grey (cyan) regions show the part of 
the parameter space where the net $h$ signal significance remains 
above $5~\sigma$. In the empty regions
between the shading and the outermost curves, the net $h$
signal drops below the $5~\sigma$ level. 
The thick grey (blue) curves indicate the 
regions where the significance of the 
$gg \to \phi \to Z^0Z^{0(*)} \to 4~\ell$ signal 
exceeds $5~\sigma$. Results are presented for 
$M_h$=120~GeV and $\lphi$= 2.5~TeV (left),
5.0~TeV (center) and 7.5~TeV (right).} 
\label{fig:compl120}
\end{figure}

\begin{figure}[h!]
\begin{center}
\begin{tabular}{c c c}
\hspace*{-0.85cm} 
\epsfig{file=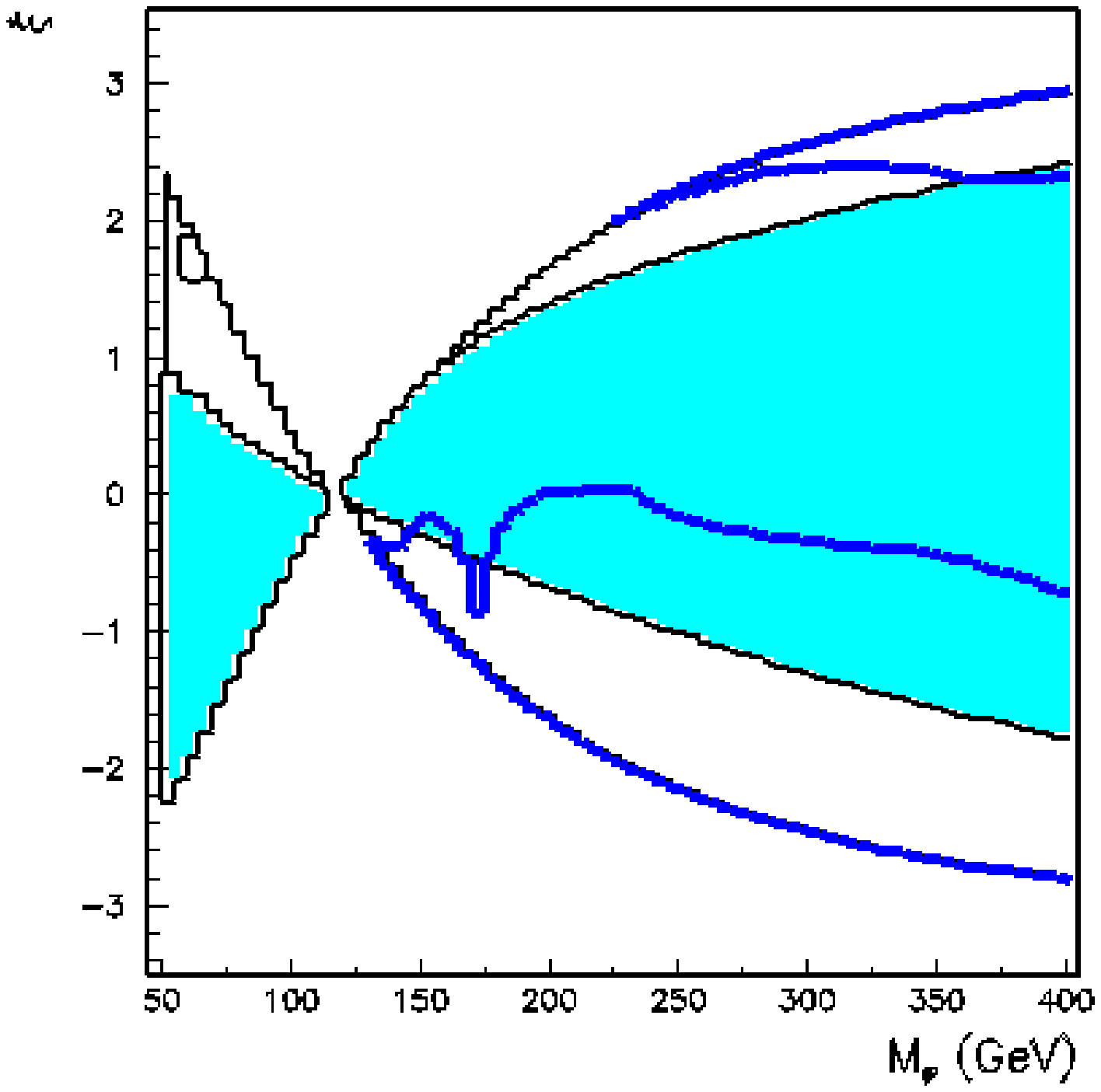,width=6.5cm,height=6.5cm} &
\hspace*{-0.7cm} 
\epsfig{file=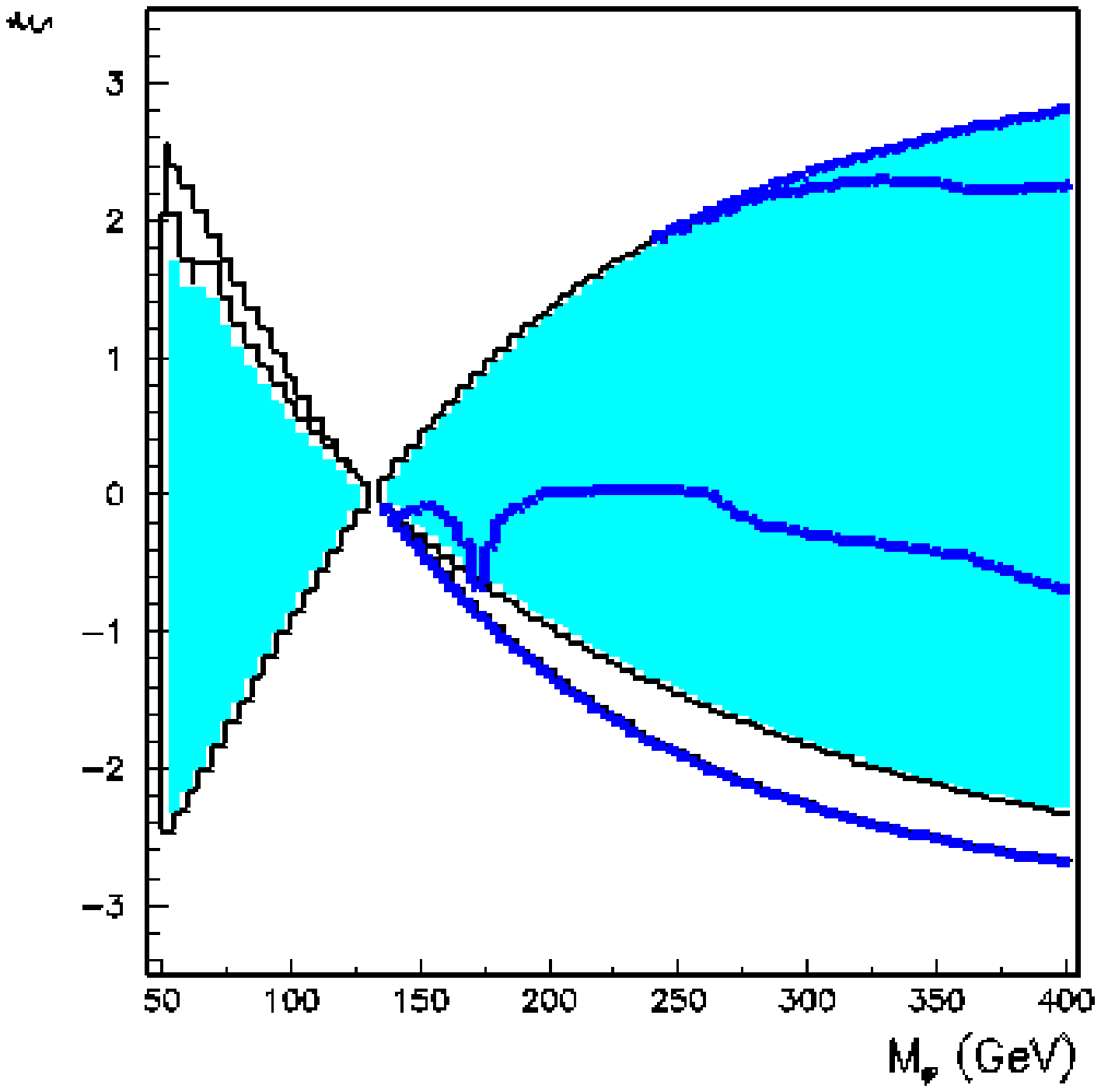,width=6.5cm,height=6.5cm} &
\hspace*{-0.7cm} 
\epsfig{file=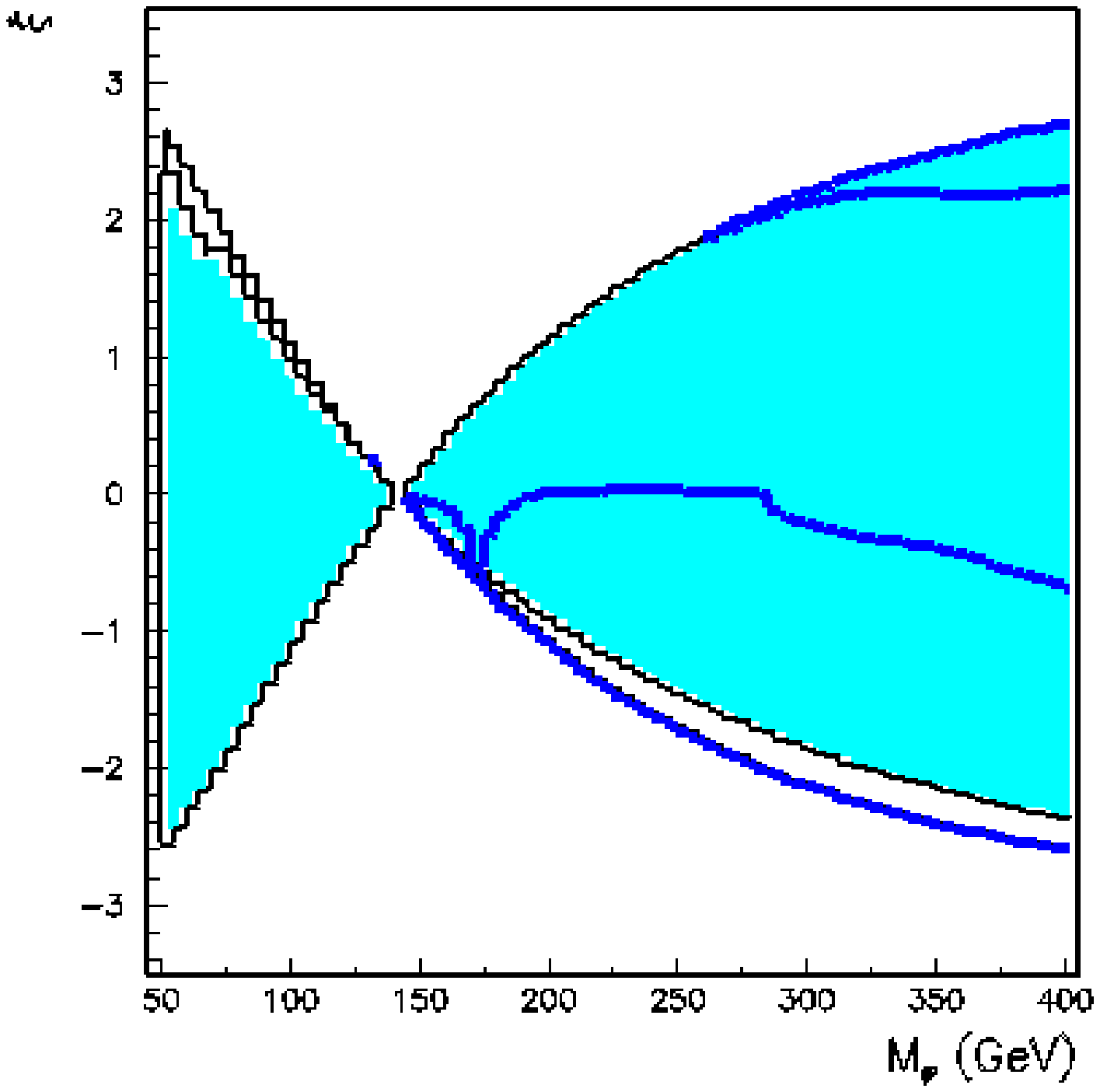,width=6.5cm,height=6.5cm} \\
\end{tabular}
\end{center}
\vspace*{-0.75cm}
\caption{\sl Same as Figure~\ref{fig:compl120} but for $M_h$ = 115~GeV (left), 
$M_h$ = 130~GeV (center) and $M_h$ = 140~GeV (right). $\lphi$ has been fixed to 
5.0~TeV.}
\label{fig:compl140}
\end{figure}

Due to the suppression, from radion mixing, of the loop-induced effective 
couplings of the $h$ (relative to the SM $H$) to 
gluon and photon pairs, the key process $gg \to h \to \gamma \gamma$ 
may fail to provide 
a significant excess over the $\gamma \gamma$ background at the LHC. 
Other modes that 
depend on the $gg$ fusion production process are suppressed too. 
For $M_\phi>M_h$, this suppression is very substantial 
for large, negative values of $\xi$. This region of significant suppression 
becomes wider at large values of $M_\phi$  and $\lphi$. 
In contrast, for $M_\phi<M_h$, the $gg\to h\to \gamma\gamma$
rate is generally only suppressed when $\xi>0$. All this is shown, in a 
quantitative way, by the contours in Figures~\ref{fig:compl120} 
and \ref{fig:compl140}.
The outermost, hourglass shaped contours define the theoretically 
allowed region. Three main regions of non-detectability may appear. 
Two are located at large values of 
$M_{\phi}$ and $|\xi|$. 
A third region appears at low $M_{\phi}$ and positive $\xi$, where
the above-noted $gg\to h\to \gamma\gamma$ suppression sets in. It
becomes further expanded when $2M_\phi<M_h$ and the decay channel
$h \to \phi \phi$ opens up, thus reducing the $h \to \gamma \gamma$ 
branching ratio. 
As shown in Figure~\ref{fig:compl140}, these regions shrink as $M_h$ increases,
since additional channels, in particular $gg\to h \to Z^0 Z^{0*}\to 4~\ell$, 
become available for Higgs discovery.

These regions are reduced by considering either 
a larger data set or $qqh$ Higgs production, in association with forward jets.
An integrated luminosity of 100~fb$^{-1}$ would remove the regions at large positive 
$\xi$ in the $\lphi=5$ and $7.5$~TeV plots of Fig.~\ref{fig:compl120}. 
Similarly, including the $qqh$, $h \to WW^* \to \ell \ell \nu \bar \nu$ channel 
in the list of the discovery modes removes the same two regions and reduces the large 
region of $h$ non-observability at negative $\xi$ values. 
 
In all these regions, a complementarity is potentially offered by the process 
$gg \to \phi \to Z^0Z^{0(*)} \to 4~\ell$, 
which becomes important for $M_{\phi} > 140$~GeV. 
At the LHC, this process would have the same event
structure as the golden SM Higgs mode 
$H \to Z^0Z^{0*} \to 4~\ell$, which has been thoroughly studied for an 
intermediate mass Higgs boson. 
By computing the  $gg \to \phi \to Z^0Z^{0(*)} \to 4~\ell$ rate relative
to that for the 
corresponding SM $H$ process and employing the LHC sensitivity curve for 
$H \to Z^0Z^{0(*)}$ of Figure~\ref{fig:sign} (left), the significance 
for the $\phi$ signal in the 
$4~\ell$ final state at the LHC can be extracted. 
Results are overlayed on 
Figures~\ref{fig:compl120} and \ref{fig:compl140}, 
assuming 30~fb$^{-1}$ of data.


Two observations are in order. 
The observability of $\phi$ production in the four lepton 
channel fills most of the gaps in $(M_h,\xi)$ parameter space
in which $h$ detection is not possible (mostly due to the suppression of 
the loop-induced  $gg\to h\to \gam\gam$ process).
The observation of at least one scalar is thus guaranteed over almost the full 
parameter phase space, with the exception of: (a) 
the region of large positive $\xi$ with $M_{\phi} <$ 140~GeV, where the 
$\phi \to Z^0Z^{0*}$ decay is phase-space-suppressed; 
and (b)  a narrow region at 
$M_{\phi} \simeq$~170~GeV due to the ramp-up of the $\phi \to W^+W^-$ channel, where a 
luminosity of order 100~fb$^{-1}$ is required to reach a $\ge 5~\sigma$ 
signal for $\phi \to Z^0Z^{0*}$. We should also note that
the $\phi \to Z^0Z^{0}$ decay is reduced for $M_\phi>2M_h$ by the 
onset of the $\phi \to hh$ decay, which can become the main decay mode. 
The resulting $hh\to b \bar{b} b \bar{b}$ topology, 
with di-jet mass constraints, may represent a viable 
signal for the LHC in its own right, but detailed studies will be needed. 
Figures~\ref{fig:compl120} and \ref{fig:compl140} also exhibit regions
of $(M_h,\xi)$ parameter space in which {\it both} the $h$ and $\phi$
mass eigenstates will be detectable.
In these regions, the LHC will observe two scalar bosons
somewhat separated in mass with the lighter (heavier) having a non-SM-like 
rate for the the $gg$-induced $\gamma\gamma$ ($Z^0Z^0$) final state.
Additional information will be required to ascertain whether these two Higgs
bosons derive from a multi-doublet or other type of extended
Higgs sector or from the present type of model with Higgs-radion mixing.

An $e^+e^-$ LC should guarantee observation of both the $h$
and the $\phi$ even in most of the regions within which detection of either
at the LHC might be difficult. Thus, this scenario
provides another illustration of the complementarity 
between the two machines in the study of the Higgs sector. 
In particular, in the region with $M_\phi>M_h$ 
the $hZ^0Z^0$ coupling is enhanced relative to the SM $HZ^0Z^0$ coupling
and $h$ detection in $e^+e^-$ collisions would be even easier
than SM $H$ detection. Further, 
assuming that $e^+e^-$ collisions could also probe 
down to $\phi Z^0Z^0$ couplings of order 
$g^2_{\phi ZZ}/g^2_{HZZ} \simeq 0.01$,  the 
$\phi$ would be seen in almost the entirety of 
the region for which $\phi$ detection at the LHC would not
be possible. In this case, the measurements of the $Z^0Z^0$ boson 
couplings of both the Higgs and the
radion particles would significantly constrain the values of the 
$\xi$ and $\lphi$ parameters of the model. 

\subsection{Determining the Nature of the Observed Scalar} 

The interplay between the emergence of the Higgs boson 
and of the radion graviscalar 
signals opens up the question of the identification of the 
nature of the newly observed particle(s).

After observing a new scalar at the LHC, some of its properties will be 
measured with sufficient accuracy to determine if they correspond to 
those expected for the SM $H$, i.e.\
for the minimal realization of the Higgs 
sector~\cite{Zeppenfeld:2000td,Zeppenfeld:qh}. 
In the presence of extra dimensions, 
further scenarios emerge. For the present discussion, we consider two 
scenarios. The first has a light Higgs boson, 
for which we take $M_h$ = 120~GeV, with 
couplings different from those predicted in the SM. 
The question here is if the anomaly
is due to an extended Higgs sector, 
such as in Supersymmetry, or rather to the mixing 
with an undetected radion. The second scenario consists of an intermediate-mass
scalar, with 180~GeV $< M <$ 300~GeV, observed alone. 
An important issue would then be the question of whether the 
observed particle is the SM-like Higgs boson or a radion, 
with the Higgs particle left undetected. This scenario is quite likely
at large negative $\xi$ and large $M_\phi$ --- 
see Figures \ref{fig:compl120} and \ref{fig:compl140}.

%\begin{figure}[hb!]
%\begin{center}
%\begin{tabular}{c c c}
%\hspace*{-0.5cm} 
%\epsfig{file=gaw_mh120_l025.eps,width=6.5cm,height=6.0cm} &
%\hspace*{-0.5cm} 
%\epsfig{file=gaw_mh120_l050.eps,width=6.5cm,height=6.0cm} &
%\hspace*{-0.5cm} 
%\epsfig{file=gaw_mh120_l075.eps,width=6.5cm,height=6.0cm} \\
%\end{tabular}
%\end{center}
%\vspace*{-0.75cm}
%\caption{\sl Ratio of couplings $g_{h\gamma\gamma}^{effective}/g_{hWW}$ normalised 
%to the SM prediction as function of $\xi$. Results are obtained for $M_h$=120~GeV and 
%$\Lambda$= 2.5~TeV (left), 5.0~TeV (center) and 7.5~TeV (right). The darker (blue) 
%curves refer to $M_{\phi}$ = 150~GeV and the lighter (red) to $M_{\phi}$ = 300~GeV.}
%\label{fig:hloop}
%\end{figure}

In the first scenario, the issue is the interpretation of discrepancies in the measured 
Higgs couplings to gauge bosons and fermions. These effects increase with $|\xi|$, 
$1/\lphi$ and $M_h/M_{\phi}$. The LHC is expected to measure some ratios of these 
couplings~\cite{Zeppenfeld:qh}. In the case of the SM $H$, the ratio 
$g_{HZZ}/g_{HWW}$ can be determined with a relative accuracy 
of 15\% to 8\% for 120~GeV $< M_H <$ 180~GeV, 
while the ratio $g_{H\tau\tau}/g_{HWW}$ and that of the 
effective coupling to photons, $g_{H\gamma\gamma}^{effective}/g_{HWW}$ can be 
determined to 6\% to 10\% for 120~GeV $< M_h <$ 150~GeV. 
Now, the Higgs-radion mixing would induce the same shifts in 
the direct couplings $g_{hWW}$, $g_{hZZ}$ and $g_{h\bar ff}$,
all being given by $d+\gamma b$ times the corresponding $H$ couplings ---
see Eq.~(\ref{couplings}).
Although this factor depends on the $\lphi$, $M_{\phi}$ and $\xi$ parameters,
ratios of couplings would remain unperturbed and correspond to those expected 
in the SM. Since the LHC measures mostly ratios of couplings,
the presence of Higgs-radion mixing could easily be missed.
One window of sensitivity to the mixing would be offered by the 
combination $g_{h\gamma\gamma}^{effective}/g_{hWW}$. 
But the mixing effects are expected
to be limited to relative variation of $\pm 5\%$ w.r.t. the SM predictions. 
Hence, the 
LHC anticipated accuracy corresponds to deviations of one unit of $\sigma$, 
or less, except for a small region at $\lphi \simeq 1$~TeV.
Larger deviations are expected for the absolute rates~\cite{Dominici:2002jv},
especially for the $gg\to h \to \gam\gam$ channel which can be dramatically
enhanced or suppressed relative to the $gg\to H \to \gam\gam$ 
prediction for larger $\xi$ values due to the large changes in
the $gg\to h$ coupling relative to the $gg\to H$ coupling. 
Of course, to detect these deviations it is necessary to control
systematic uncertainties for the absolute $\gam\gam$ rate. 
All the above remarks would also apply to distinguishing between the light 
Higgs of supersymmetry, which would be SM-like
assuming an approximate decoupling limit, and the $h$ of the Higgs-radion
scenario. In a non-decoupling two-doublet model, the light Higgs
couplings to up-type and 
down-type fermions can be modified differently with respect to those of
the SM $H$, and LHC measurements of coupling ratios
would detect this difference.

\begin{figure}
\begin{center}
\begin{tabular}{c c}
\hspace*{-0.75cm} 
\epsfig{file=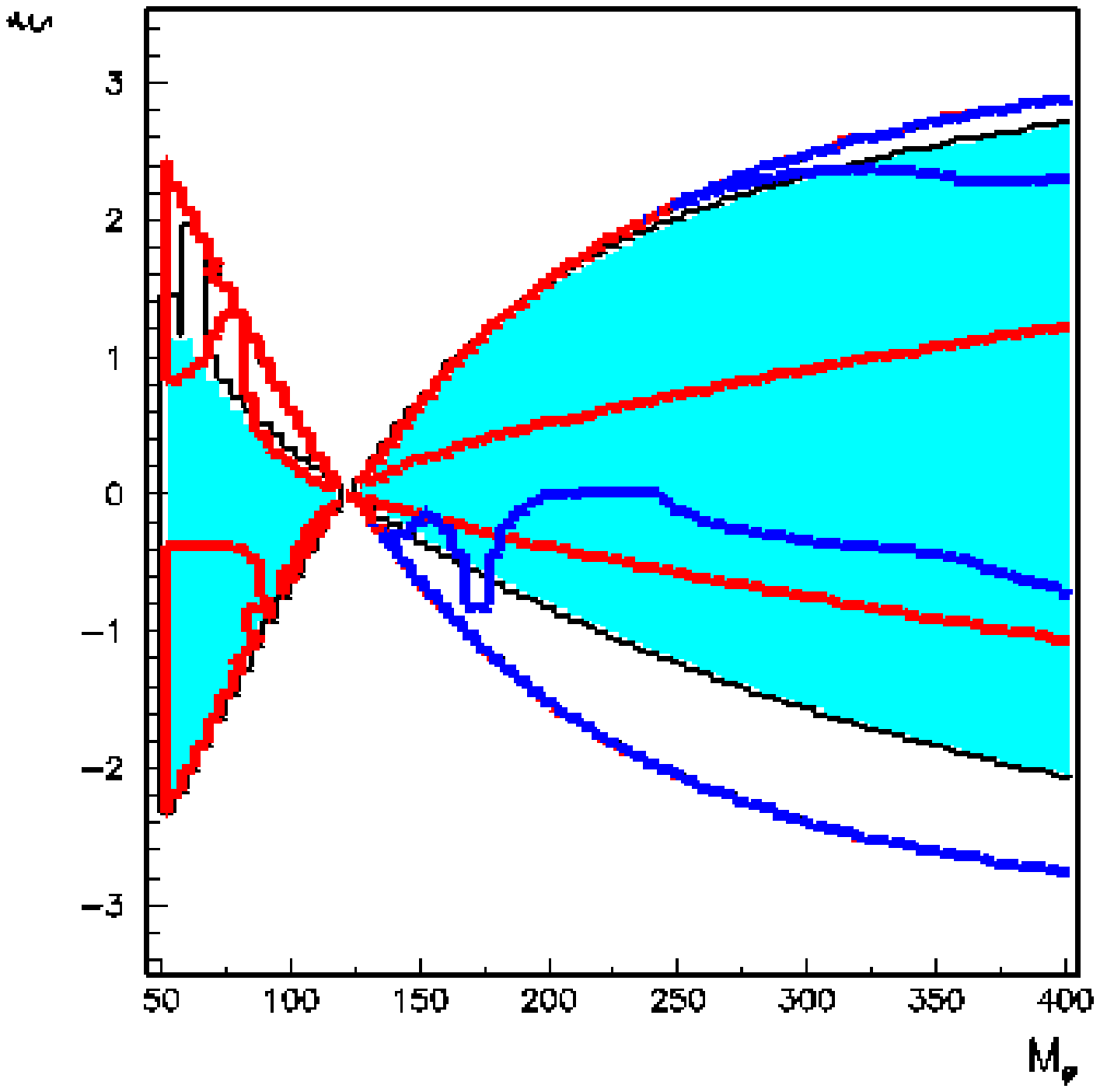,width=8.5cm,height=8.0cm} &
\epsfig{file=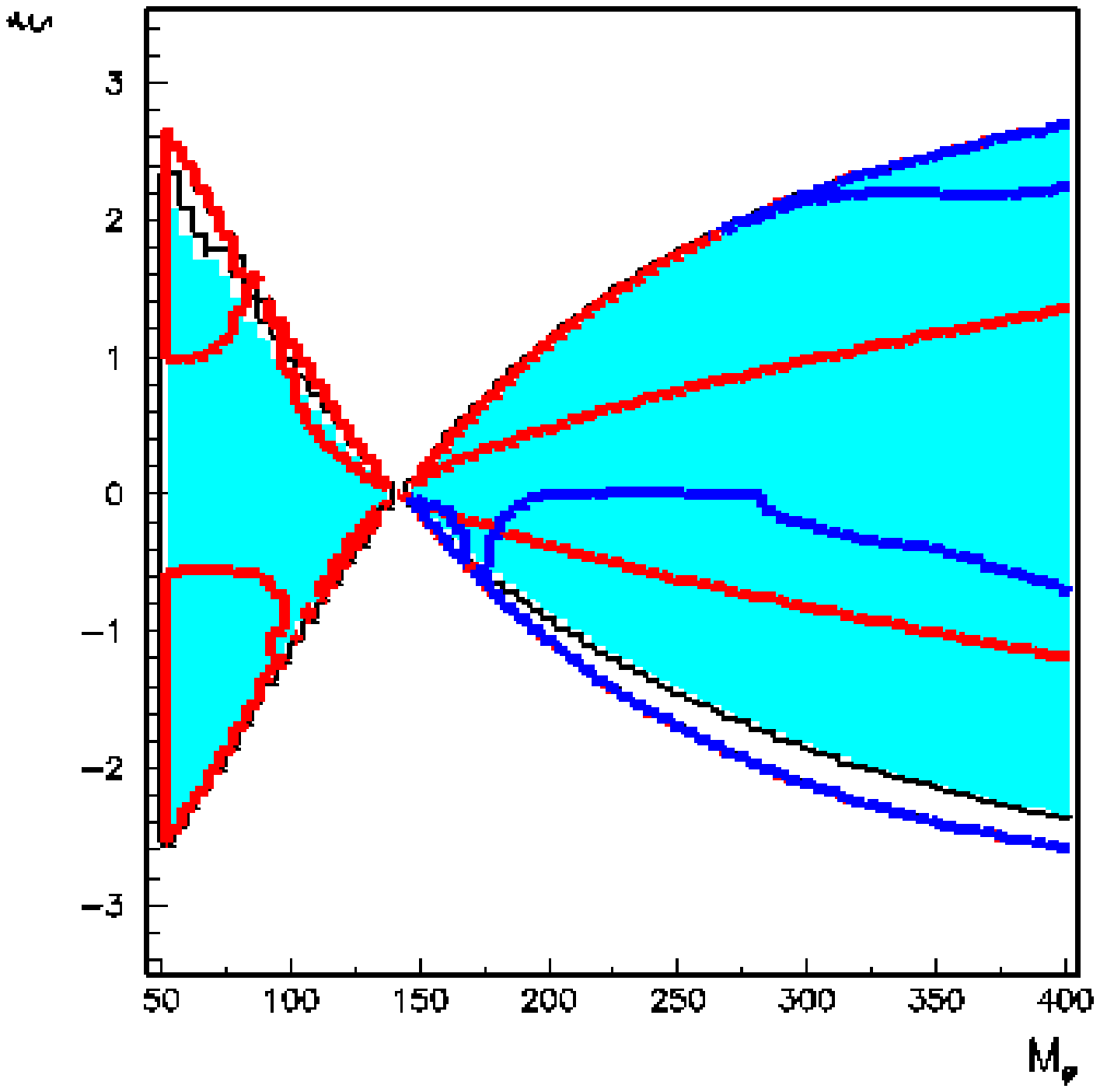,width=8.5cm,height=8.0cm} \\
\end{tabular}
\end{center}
\caption{\sl Same as Figures~\ref{fig:compl120}
and \ref{fig:compl140} for $M_h$ = 120~GeV (left), 
140~GeV (right) and $\lphi$ = 5~TeV with added contours, 
indicated by the medium 
grey (red) curves, showing 
the regions where the LC measurements of the $h$ couplings to 
$b \bar{b}$ and $W^+W^-$ would provide a $>2.5~\sigma$ 
evidence for the radion mixing effect. (Note: the grey (red) lines
are always present along the outer edge of the hourglass in the $M_\phi>M_h$
region, but are sometimes buried under the darker (blue) curves.
In this region, the $>2.5~\sigma$ regions lie between the outer hourglass
edges and the inner grey (red) curves.)}
\label{fig:lc}
\end{figure}

A TeV-class LC has the capability of extending the coupling
measurements to all fermions separately with accuracies of 
order 1\%-5\% and achieves a 
determination of the total width to 4\% - 6\% accuracy~\cite{hlc}. 
This is important for the scenario we propose since it would provide 
enough measurements 
and sufficient accuracy to detect Higgs-radion  mixing for 
moderate to large $\xi$ values~\cite{Rizzo:2002za}. This 
is shown in Figure~\ref{fig:lc} 
by the additional contours, which indicate the regions 
where the discrepancy with the SM predictions 
for the Higgs couplings to pairs of $b$ 
quarks and $W$ bosons exceeds 2.5~$\sigma$.
In particular, the {\it combination} of the direct observation of 
$\phi \to Z^0Z^{0*}$ at the 
LHC and the precision measurements of the Higgs properties at a $e^+e^-$ LC 
will extend our ability to distinguish between the Higgs-radion mixing
scenario and the SM $H$ scenario 
to a large portion of the regions where at the LHC
only the $h$ or only the $\phi$ is detected and determining
that the observed boson is not the SM $H$ is difficult. 
Further, close to the edges of the hourglass-shaped 
allowed region, the LC will also be able to detect 
$\phi$ production directly through the 
process $e^+e^- \to Z^0 \phi$. In particular, this process will guarantee the 
observability of the $\phi$ in the low $M_{\phi}$ region, 
which is most difficult for the LHC.

\begin{figure}
\begin{center}
\begin{tabular}{c c}
\hspace*{-0.75cm} 
\epsfig{file=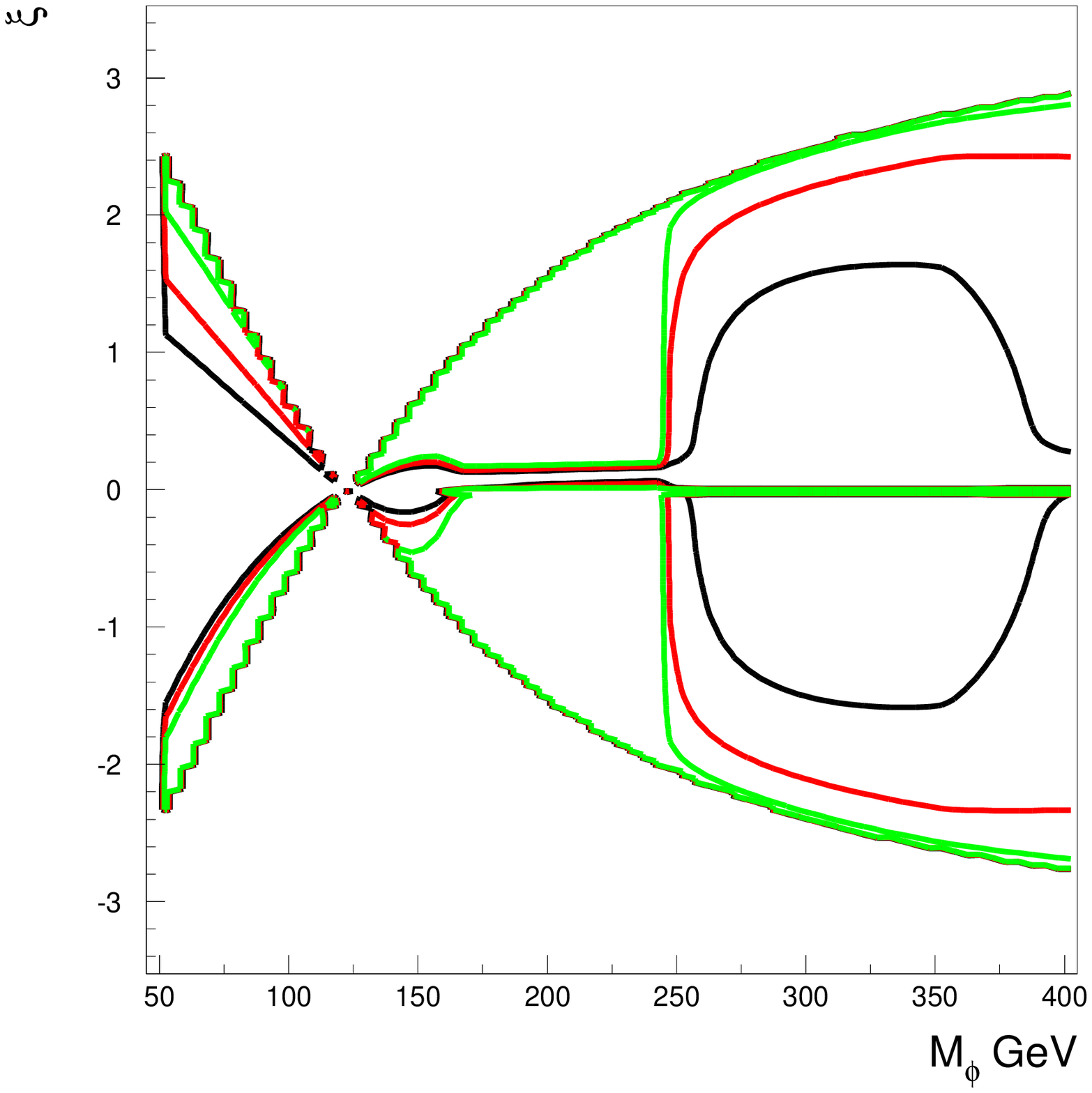,width=8.5cm,height=8.0cm} &
\epsfig{file=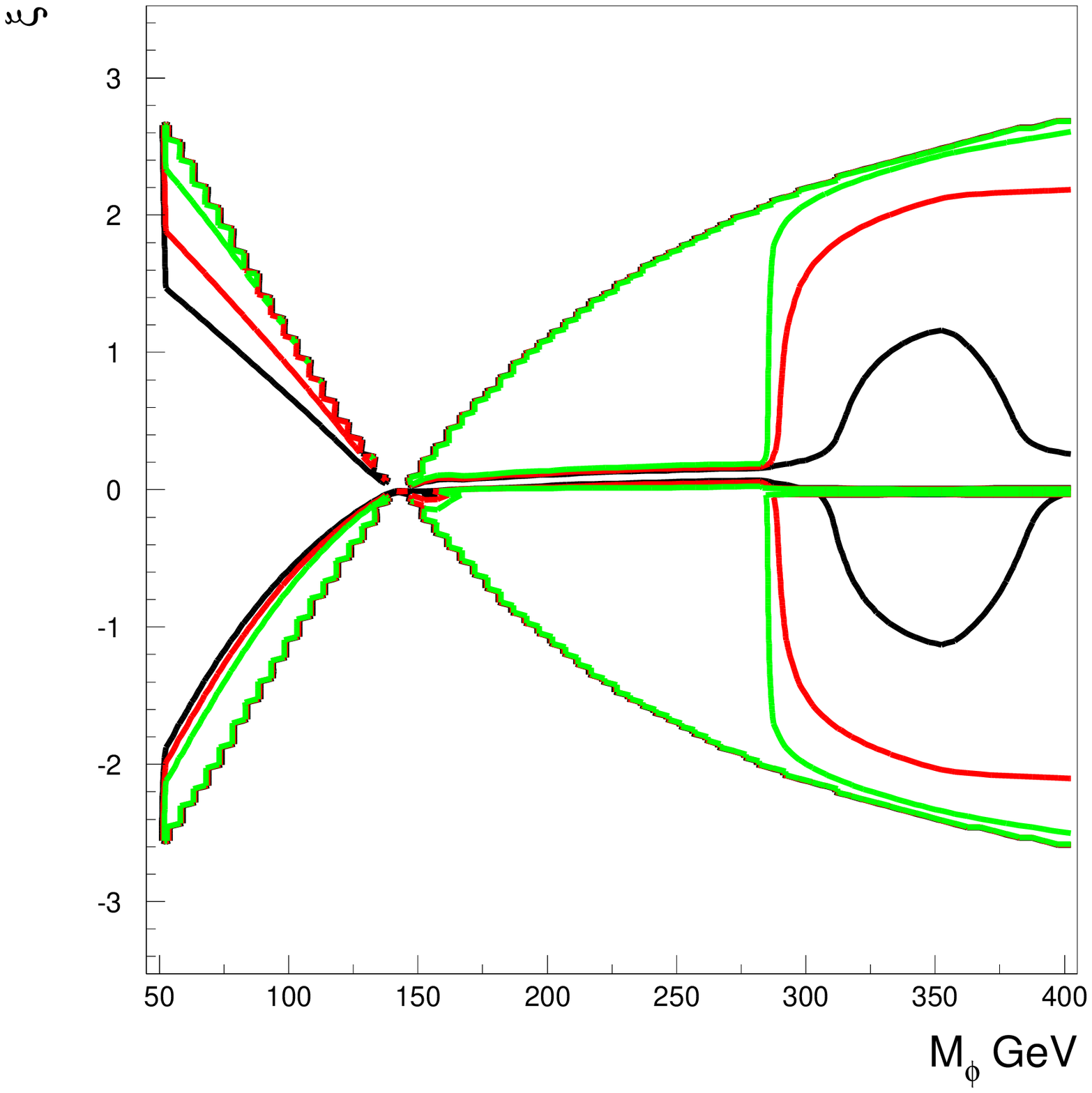,width=8.5cm,height=8.0cm} \\
\end{tabular}
\end{center}
\vspace*{-0.75cm}
\caption{\sl Ratio $\frac{{\mathrm{BR}}(\phi \to Z^0Z^{0(*)})}{{\mathrm{BR}}(H_{SM} \to 
Z^0Z^{0(*)})}$ as function of $M_{\phi}$ and $\xi$. The curves indicate the 0.7 (black), 
0.8 (medium grey/red) and 0.9 (light grey/green) contours. Results are obtained for 
$M_h$=120~GeV (left) and $M_h$ = 140~GeV (right) and $\lphi$=5.0~TeV. The radion can 
be distinguished from the SM Higgs particle at intermediate values of its mass, past the 
threshold for the $\phi \to hh$ decay.}
\label{fig:brzz120}
\end{figure}

If, at the LHC, an intermediate mass scalar is observed alone, 
its non-SM-like nature can, in some cases, be determined 
through measurement of its production yield and its couplings. In 
particular, in the region at large, 
negative $\xi$ values where $\phi$ production is visible
whereas $h$ production is not, the yield of $Z^0Z^0 \to 4~\ell$ from 
$\phi$ decay can differ by a factor of 2 or more from that expected
for a SM $H$ (depending upon the value of $M_\phi$ --- see Figure~13
of Ref.~~\cite{Dominici:2002jv}). 
For $M_\phi$ such that  $\phi\to hh$ decays are
not allowed, the deviations arise from the substantial
differences between the $gg\to \phi$ coupling and the $gg\to H$
coupling. For $M_\phi>2M_h$ this rate is also sensitive to the
exclusive branching fraction. Figure~\ref{fig:brzz120} shows the 
ratio of the $Z^0Z^{0(*)}$ decay branching 
fraction for the radion to that for the SM $H$. The figure shows
that branching ratio differences
are expected to be below 10\% for radions with mass up to twice
the Higgs mass. Such a small difference would not have a big
impact compared to the possibly large
deviations of $gg\to h/gg\to H$ relative to unity.
However, past the threshold for $\phi \to hh$ decays, the $Z^0Z^0$ branching 
fraction is significantly affected 
away from $\xi \simeq 0$. The combination of a 
reduced $Z^0Z^0 \to 4~\ell$ rate and the possibility 
to observe $\phi \to hh$ decays, 
ensures that the LHC could positively identify the existence of 
the radion in the 
region $M_{\phi} >2M_h$, $\xi \ne 0$. 

To conclude, we should note
that the Higgs-radion sector is not the only means for probing
the Randall-Sundrum type of model.  The scenarios 
considered here will also yield the distinctive signature 
of KK graviton excitation production at the LHC~\cite{Davoudiasl:2000wi}. 
This easily observed signal will serve as a warning to look for
a possibly mixed Higgs-radion sector.

\section{Conclusion}

Perspectives for light Higgs searches at the LHC have been reviewed 
for models with warped extra dimensions, which introduce the radion graviscalar. 
The mixing of the Higgs field with the radion field induces changes in 
the production and decay properties of the Higgs boson mass eigenstate.
Such changes may weaken, or even invalidate, the expectations obtained in earlier 
studies for observability of the Higgs boson. However, for almost the entire region 
of the parameter phase space where the suppression of the Higgs signal yield 
causes the overall signal significance at the LHC to drop below 5~$\sigma$, the 
radion eigenstate $\phi$ can be observed in the $gg \to \phi \to Z^0Z^{0(*)} \to 4~\ell$ 
process instead. An $e^+e^-$ linear collider would effectively complement the LHC both 
for the Higgs observability, including the most difficult region at low $M_{\phi}$ 
and positive $\xi$ values, and for the detection of the radion mixing effects, 
through the precision measurements of the  Higgs particle couplings to various types of 
particle pairs. 

\vspace*{0.75cm}
\noindent
We wish to thank F.~Gianotti, B. Grzadkowski, J.~Hewett, A.~Nikitenko, T.~Rizzo
and M. Toharia, for discussions 
and suggestions. JFG is supported by the U.S. Department of Energy
and the Davis Institute for High Energy Physics.


\begin{thebibliography}{99}
\itemsep=0in
\bibitem{rs}
L. Randall, R. Sundrum,
%``A large mass hierarchy from a small extra dimension'',
Phys. Rev. Lett. {\bf 83} (1999) 3370, [arXiv:hep-ph/9905221];\\
L. Randall, R. Sundrum,
%``An alternative to compactification'',
Phys. Rev. Lett. {\bf 83} (1999) 4690, [arXiv:hep-th/9906064].
%

%\cite{Bae:2000pk}
\bibitem{Bae:2000pk}
S.~B.~Bae, P.~Ko, H.~S.~Lee and J.~Lee,
%``Phenomenology of the radion in the Randall-Sundrum scenario at  colliders,''
Phys.\ Lett.\ B {\bf 487} (2000) 299
[arXiv:hep-ph/0002224].
%%CITATION = HEP-PH 0002224;%%

%\cite{Davoudiasl:1999jd}
\bibitem{Davoudiasl:1999jd}
H.~Davoudiasl, J.~L.~Hewett and T.~G.~Rizzo,
%``Phenomenology of the Randall-Sundrum gauge hierarchy model,''
Phys.\ Rev.\ Lett.\  {\bf 84} (2000) 2080
[arXiv:hep-ph/9909255].
%%CITATION = HEP-PH 9909255;%%

%\cite{Cheung:2000rw}
\bibitem{Cheung:2000rw}
K.~Cheung,
%``Phenomenology of radion in Randall-Sundrum scenario,''
Phys.\ Rev.\ D {\bf 63} (2001) 056007
[arXiv:hep-ph/0009232].
%%CITATION = HEP-PH 0009232;%%

%\cite{Davoudiasl:2000wi}
\bibitem{Davoudiasl:2000wi}
H.~Davoudiasl, J.~L.~Hewett and T.~G.~Rizzo,
%``Experimental probes of localized gravity: On and off the wall,''
Phys.\ Rev.\ D {\bf 63} (2001) 075004 
[arXiv:hep-ph/0006041].
%%CITATION = HEP-PH 0006041;%%

%\cite{Park:2000xp}
\bibitem{Park:2000xp}
S.~C.~Park, H.~S.~Song and J.~Song,
%``Radion effects on the production of an intermediate-mass scalar and Z  at
%LEP II,''
Phys.\ Rev.\ D {\bf 63} (2001) 077701
[arXiv:hep-ph/0009245].
%%CITATION = HEP-PH 0009245;%%


\bibitem{wells_mix}
G. Giudice, R. Rattazzi, J. Wells, 
%``Graviscalars from higher-dimensional metrics 
%and curvature-Higgs mixing'', 
Nucl. Phys. B {\bf 595} (2001) 250  [arXiv:hep-ph/0002178]. 
%

\bibitem{csaki_mix}
C. Csaki, M.L. Graesser, G.D. Kribs, 
%``Radion dynamics and electroweak physics'',
Phys. Rev. D {\bf 63} (2001) 065002-1 [arXiv:hep-th/0008151].
%

%\cite{Han:2001xs}
\bibitem{Han:2001xs}
T.~Han, G.~D.~Kribs and B.~McElrath,
%``Radion effects on unitarity in gauge-boson scattering,''
Phys.\ Rev.\ D {\bf 64} (2001) 076003
[arXiv:hep-ph/0104074].
%%CITATION = HEP-PH 0104074;%%

%\cite{Chaichian:2001rq}
\bibitem{Chaichian:2001rq}
M.~Chaichian, A.~Datta, K.~Huitu and Z.~h.~Yu,
%``Radion and Higgs mixing at the LHC,''
Phys.\ Lett.\ B {\bf 524} (2002) 161
[arXiv:hep-ph/0110035].
%%CITATION = HEP-PH 0110035;%%

\bibitem{Azuelos:fv}
G.~Azuelos, D.~Cavalli, H.~Przysiezniak and L.~Vacavant,
%``Search For The Radion Using The Atlas Detector,''
Eur.\ Phys.\ J.\ direct C {\bf 4} (2002) 16.
%%CITATION = EPHJD,C4,16;%%

\bibitem{Hewett:2002nk}
J.~L.~Hewett and T.~G.~Rizzo,
%``Shifts in the properties of the Higgs boson from radion mixing,''
arXiv:hep-ph/0202155.
%%CITATION = HEP-PH 0202155;%%

\bibitem{Dominici:2002jv}
D.~Dominici, B.~Grzadkowski, J.~F.~Gunion and M.~Toharia,
%``The scalar sector of the Randall-Sundrum model,''
arXiv:hep-ph/0206192.
%%CITATION = HEP-PH 0206192;%%

\bibitem{johum}
J.J. van der Bij, Acta Physica Polonica, B {\bf 25} (1994) 827; 
R.~Raczka, M.~Pawlowski, Found. Phys. 24 (1994) 1305, [arXiv:hep-th/9407137].

\bibitem{atlas-tdr}
ATLAS Collaboration, {\sl Detector and Physics Performance - Technical Design Report}, 
LHCC 99-14/15.

\bibitem{cmsnote}
A. Abdullin {\it et al.}, {\it Summary of the CMS Discovery Potential for the Higgs
boson}, CMS Note in preparation.

\bibitem{Djouadi:1997yw}
A.~Djouadi, J.~Kalinowski and M.~Spira,
%``HDECAY: A program for Higgs boson decays in the standard model and its  supersymmetric extension,''
Comput.\ Phys.\ Commun.\  {\bf 108} (1998) 56, 
[arXiv:hep-ph/9704448].
%%CITATION = HEP-PH 9704448;%%

\bibitem{Zeppenfeld:2000td}
D.~Zeppenfeld, R.~Kinnunen, A.~Nikitenko and E.~Richter-Was,
%``Measuring Higgs boson couplings at the LHC,''
Phys.\ Rev.\ D {\bf 62} (2000) 013009, 
[arXiv:hep-ph/0002036].
%%CITATION = HEP-PH 0002036;%%

\bibitem{Zeppenfeld:qh}
D.~Zeppenfeld,
%``Higgs Physics At The Lhc,''
Int.\ J.\ Mod.\ Phys.\ A {\bf 16S1B} (2001) 831.
%%CITATION = IMPAE,A16S1B,831;%%

\bibitem{hlc}
M.~Battaglia and K.~Desch, Proc. of the 5$^{th}$ {\sl Int. Linear Collider Workshop 
LCWS~2000}, Fermilab, Batavia (IL), USA, AIP Conf.\ Proc.\ {\bf 578}, 163, 
[arXiv:hep-ph/0101165].

\bibitem{Rizzo:2002za}
T.~G.~Rizzo,
%``Effects of radion mixing on the standard model Higgs boson,''
arXiv:hep-ph/0209076.
%%CITATION = HEP-PH 0209076;%%

\end{thebibliography}
\end{document}